\documentclass[conference]{IEEEtran}%
\IEEEoverridecommandlockouts
\usepackage{amsfonts}
\usepackage{amsmath}
\usepackage{balance}
\usepackage{cite}
\usepackage{subfigure}
\PassOptionsToPackage{hyphens}{url}
\usepackage{url}

\usepackage[pdftex]{graphicx}
\usepackage{epstopdf}
\usepackage{multirow}

\usepackage{pifont}
\newtheorem{lemma}{Lemma}

\newcommand{\pt}{p_{\mathsf{t}}}
\newcommand{\thetat}{\theta_{\mathsf{t}}}
\newcommand{\thetar}{\theta_{\mathsf{r}}}
\newcommand{\Gt}{G_{\mathsf{t}}}
\newcommand{\gt}{g_{\mathsf{t}}}
\newcommand{\Gr}{G_{\mathsf{r}}}
\newcommand{\gr}{g_{\mathsf{r}}}
\newcommand{\alphaL}{\alpha_{\mathsf{L}}}
\newcommand{\alphaN}{\alpha_{\mathsf{N}}}

\newcommand{\IPhi}{I_{\mathbf{\Phi}}}
\newcommand{\BPhi}{\mathbf{\Phi}}
\newcommand{\RB}{R_{\mathsf{LOS}}}
\newcommand{\pbr}{p_{\mathsf{b}}(r)}
\newcommand{\rout}{r_{\mathsf{net}}}

\newcommand{\be}{\begin{eqnarray}}
\newcommand{\ee}{\end{eqnarray}}

\usepackage[usenames]{color}

\begin{document}
\hyphenation{multi-symbol}
\title{Analysis of Millimeter Wave Networked Wearables in Crowded Environments}
\author{ Kiran Venugopal, Matthew C. Valenti, and Robert W. Heath, Jr. \\ 
\thanks{This work was supported in part by the Intel-Verizon 5G research program and the National Science Foundation under Grant No. NSF-CCF-1319556. Kiran Venugopal and Robert W. Heath, Jr. are with the University of Texas, Austin, TX, USA. Matthew C. Valenti is with West Virginia University, Morgantown, WV, USA. Email: \tt{kiranv@utexas.edu, valenti@ieee.org, rheath@utexas.edu}}
}
\date{}
\maketitle

\vspace{-1cm}
\thispagestyle{empty}
\begin{abstract}
The millimeter wave (mmWave) band has the potential to provide high throughput among wearable devices. When mmWave wearable networks are used in crowded environments, such as on a bus or train, antenna directivity and orientation hold the key to achieving Gbps rates. Previous work using stochastic geometry often assumes an infinite number of interfering nodes drawn from a Poisson Point Process (PPP). Since indoor wearable networks will be isolated due to walls, a network with a finite number of nodes may be a more suitable model. In this paper, we characterize the significant sources of interference and develop closed-form expressions for the spatially averaged performance of a typical user's wearable communication link. The effect of human body blockage on the mmWave signals and the role of network density are investigated to show that an increase in interferer density reduces the mean number of significant interferers.
\end{abstract}

\section{Introduction}
In a mobile wearable communication network, several devices positioned around a user's body may communicate with each other in a peer-to-peer fashion or client-server fashion \cite{Starner2014}. Unlike conventional wireless body area networks (WBANs), a wearable communication network may also include high-end devices like augmented reality glasses and 3D displays that require data rates of the order of giga bits per second, in addition to low-rate devices like health monitoring gadgets. While a controlling node like the user's smartphone may coordinate the communication between the different devices attached to the given user, the wearable networks associated with different users will likely be uncoordinated. This is problematic when supporting high data rate devices in crowded environments, where interference from other-user wearable networks is high \cite{Winter:2015}. 

Prior work on interference modeling for WBANs have mostly considered sub $10~ \text{GHz}$ frequency bands \cite{Kazemi:2010, Wen-Bin:2011, Cotton:CABAN}. Even with one active transmitting device per user, supporting multiple uncoordinated wearable communication networks becomes unfeasible at high user densities like $1-2~\text{users/m}^{-2}$ due to resource scarcity \cite{Winter:2015}. This motivates us to consider the millimeter wave (mmWave) band as a potential candidate for wearable networks as it can support high data rates with reasonable isolation and directivity features \cite{mmWaveBook, Cotton:60_2.5GHzcompare}. The unique propagation characteristics of mmWave frequencies imply that both human bodies and buildings act as blockages for signals. Human bodies that can result in 20-40 dB of attenuation for mmWave signals \cite{Lu:ZTE,Bai:Asilomar14} are the main source of blockage in the mmWave wearable networks context. This was considered in \cite{mmWave:2015} for computing the exact signal-to-interference-plus-noise-ratio (SINR) distribution for a given fixed location of mmWave wearable networks, leveraging tools from \cite{torrieri:2012} and assuming the fading is Nakagami distributed.

In this paper, we analyze the performance of mmWave based wearable networks modeling human bodies as the main source of blockages and develop an analytic approach for computing spatially averaged SINR performance as seen by a typical user. We assume the user density is very high and the wearable networks across users are uncoordinated. In this scenario, the users act as both sources of blockages and also carry potentially interfering devices with respect to a typical user's wearable communication link. Our stochastic geometry based analysis includes this aspect and system parameters like antenna gain and beam-width, network geometry, and user density. We show that the implication of higher user density is very different for the wearables scenario compared to mmWave cellular systems. Using results from random shape theory, we categorize interferers as \textit{strong} and \textit{weak}; the average interference due to the weak interferes is assumed to add to the thermal noise. Our approach mitigates the high computational complexity required to compute outage probability in \cite{mmWave:2015} for larger Nakagami coefficients and for the limiting case of an AWGN channel, in particular.
\begin{figure}[h]
\centering
\includegraphics[width=2.2in]{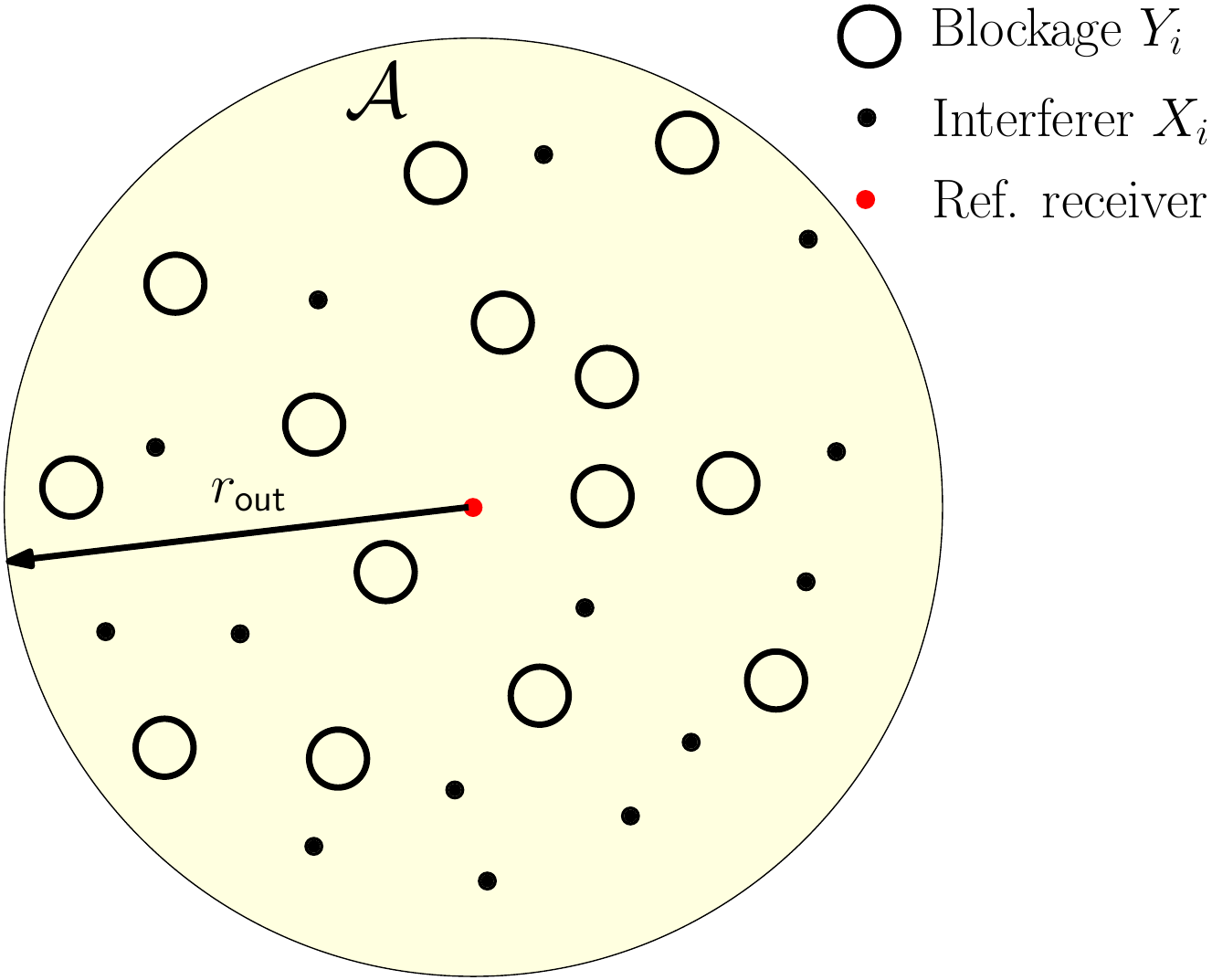}
\vspace{-0.1in}
\caption{The network region $\mathcal{A}$ with the reference receiver at the center and potentially interfering transmitters $X_i$ and blockages $Y_i$ drawn from independent point processes}     
\label{fig:network_region}        
\end{figure}

\section{System Model}
We consider a circular network region $\mathcal{A}$ with radius $\rout$ consisting of a reference transmitter-receiver pair whose location is relatively fixed, and potentially interfering transmitters that are drawn from a non-homogeneous Poisson Point Process (PPP) $\BPhi$ that has intensity $\lambda$ inside $\mathcal{A}$, and zero outside of it (see Fig. \ref{fig:network_region}). A circular network region with the reference receiver at the center captures the location of interferers relative to a reference receiver's perspective. We represent the transmitters and their locations as a complex number $X_i = R_i e^{j \phi_i}$, where $R_i = |X_i|$ is the distance of the $i^{th}$ transmitter from the reference receiver and $\phi_i = \angle X_i$ is the corresponding angle to $X_i$.

The human body blockages are modeled  as circles of diameter $W$. The location of their centers are denoted by $Y_i$ and are drawn from a density $\lambda$ PPP that is independent\footnote{In reality, the locations of the blockages and interferers are correlated (since humans hold the devices and serve as blockages), but the effect of the spatial correlation is negligible (per our work in \cite{mmWave:2015}).} of $\BPhi$. A transmitter $X_i$ is said to be blocked if the path from $X_i$ to the reference receiver intersects any circle centered at $Y_j$ or if $X_i$ falls within a diameter-$W$ circle of some blockage $Y_j$.

We assume the power gain $h_i$ due to fading for the wireless link from $X_i$ to the reference receiver is a normalized Gamma distributed random variable with parameter $m_i$. If $X_i$ is blocked, we say the communication is \textit{line of sight} (LOS) and $m_i = m$; otherwise, we say that it is \textit{non-LOS} (NLOS). The path-loss exponent for the link from $X_i$ is denoted $\alpha_i$ which takes value $\alphaL$ if $X_i$ is LOS and $\alphaN$ if it is NLOS. The link between the reference transmitter-receiver pair is assumed to be LOS. We further assume that with probability $\pt$ each $X_i$ transmits with power $\mathsf{P}$.

As considered before in \cite{mmWave:2015,mmWave_arXiv:2015}, we assume a sectorized antenna array pattern at the transmitter and the receiver antennas that can be characterized by three parameters - the beamwidth of the antenna main-lobe $\theta$, the main-lobe gain $G$ within beamwidth $\theta$, and the side-lobe gain $g$ outside the main-lobe. We use subscripts $\mathsf{t}$ and $\mathsf{r}$ to denote the antenna parameters at the transmitter and the receiver, respectively. The reference receiver is assumed to be pointed in the direction $\phi_0$ where the reference transmitter is assumed to be located.

\section{Interference Analysis}
\begin{figure}
\centering
\includegraphics[totalheight = 1.9in,width=1.9in]{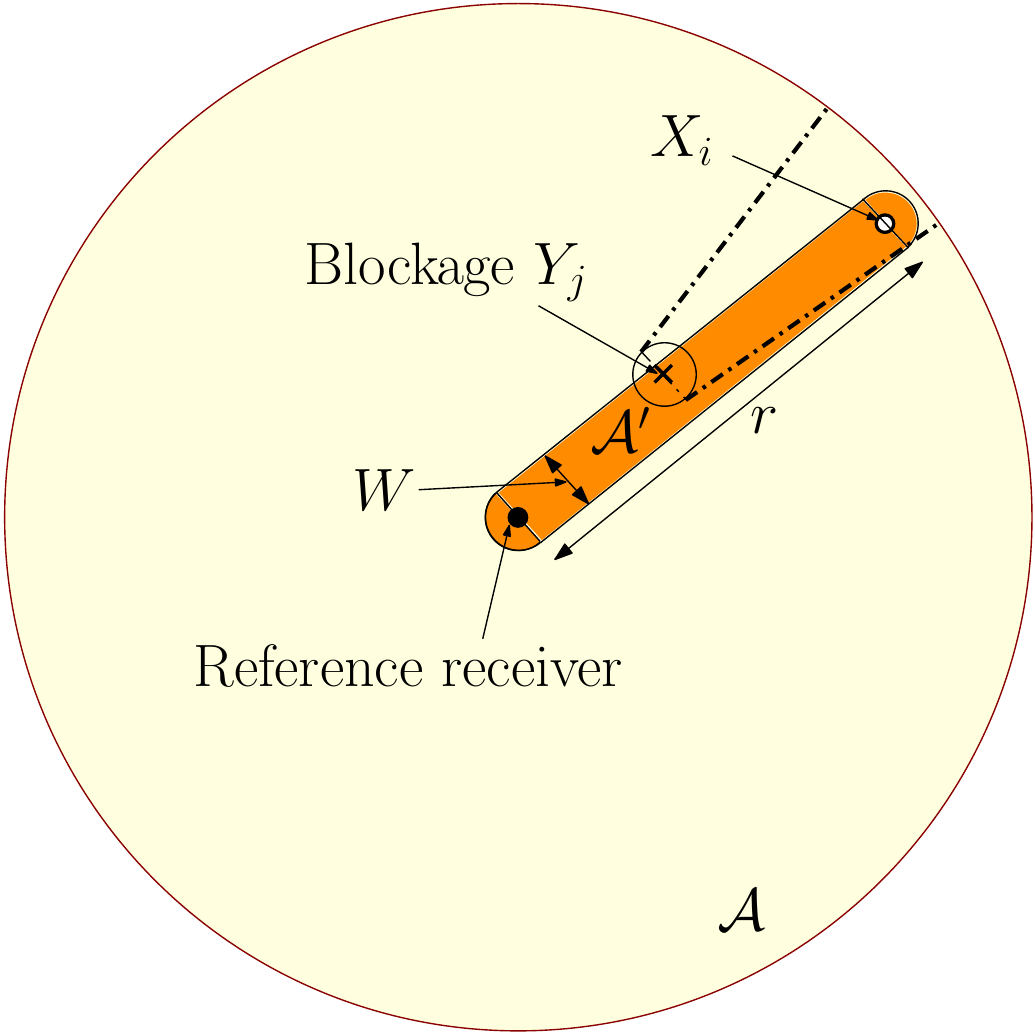}
\vspace{-0.1in}
\caption{The blocking region for user $X_i$. A blockage $Y_j$ that lies within this blocking region and its \textit{blocking wedge} are also marked for illustration.}     
\label{fig:blocking_region}        
\end{figure}

We define the blockage probability $\pbr$ as the probability that an interferer at distance $r$ from the reference receiver is blocked. We evaluate $\pbr$ as given in the following Lemma. 
\begin{lemma}
\label{lemma:snowboard}
When blockages have diameter $W$, the probability that an interferer at distance $r$ from the reference receiver is blocked  is 
\begin{eqnarray}
\pbr =   1 - \exp(-\lambda |{\mathcal{A}}'|), ~ \mathrm{where} ~ |{\mathcal{A}}'| = rW + \pi \frac{W^2}{4}. \label{Equation:block_prob}
\end{eqnarray}
\end{lemma} 
This can be derived by noting that an interferer at distance $r$ from the reference is blocked whenever there is a blockage in the blocking region $\mathcal{A}'$ as shown in Fig. \ref{fig:blocking_region}. Since the blockages are drawn from a PPP of density $\lambda$, Lemma \ref{lemma:snowboard} follows.

\begin{figure}
\centering
\includegraphics[totalheight=2.6in, width=2.8in]{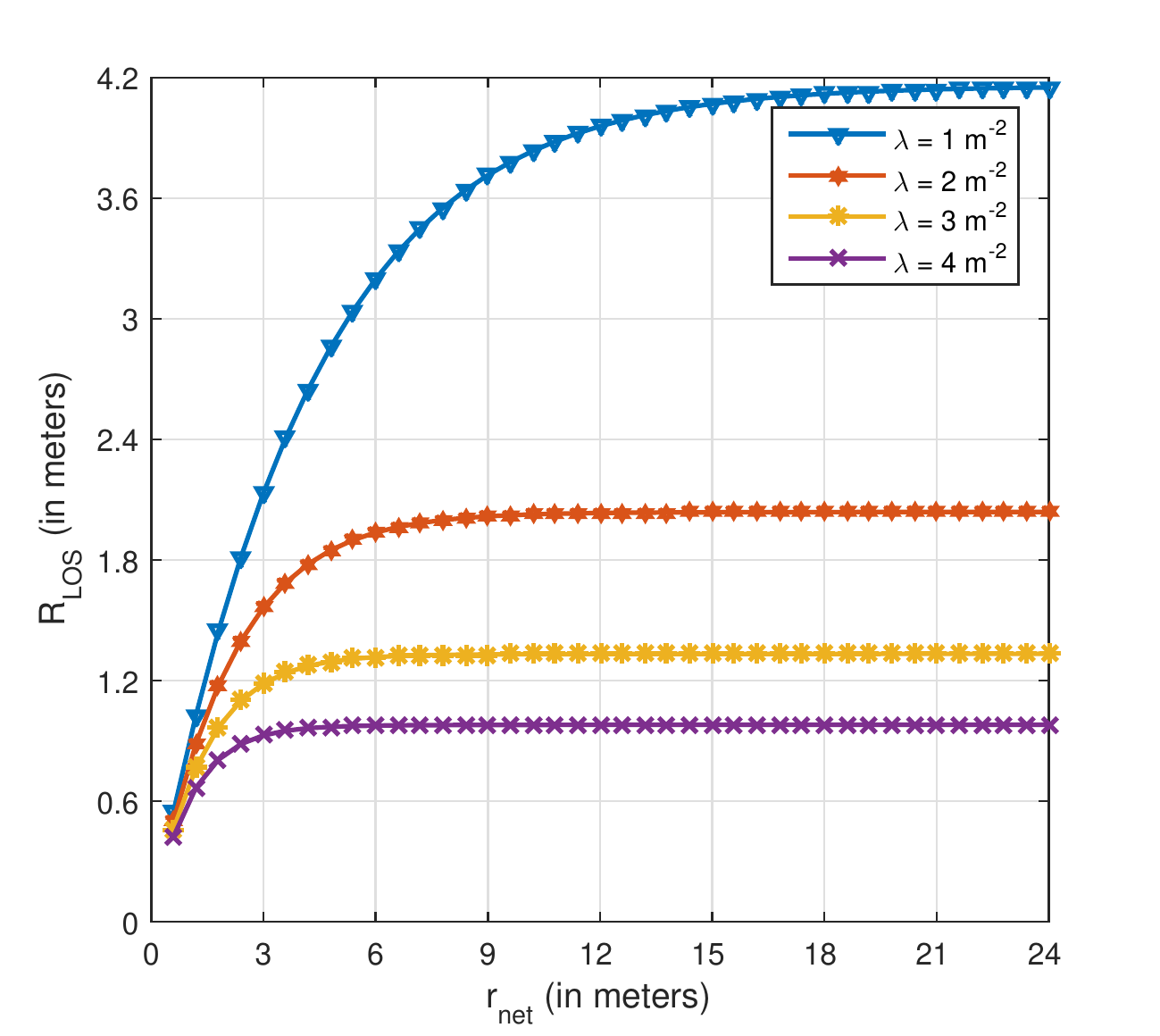}
\vspace{-0.1in}
\caption{Plot showing the variation in the LOS ball radius with respect to the network radius for different values of interferer densities.}     
\label{fig:RoutVsRB}        
\end{figure}

To simplify analysis, we define an equivalent ball of radius $\RB$ denoted as $\mathcal{B}(0, \RB)$ to replace the irregular and random LOS boundary.  The value of $\RB$ is evaluated by matching the first moments (Criterion 1 of \cite{bai:2015}). With the distance dependent blockage probability $\pbr$ as given in \eqref{Equation:block_prob}, the mean number of non-blocked interferers ${\mathbb{E}}({\mathbf{I}}_{\mathsf{LOS}})$ is
\be
{\mathbb{E}}({\mathbf{I}}_{\mathsf{LOS}}) &=& \hspace{-0.1in}2\pi \lambda \int_0^{\rout}(1-\pbr)r \mathsf{d}r\\
&=& \hspace{-0.1in}\frac{2\pi e^{\frac{-\lambda \pi W^2}{4}}}{W^2 \lambda}\left( 1- e^{-\lambda W \rout}\left(1 + \lambda W \rout \right)\right). \label{Equation:MeanLOSint}
\ee
Equating \eqref{Equation:MeanLOSint} to the mean number of interferers in the equivalent LOS ball of radius $\RB$, we find the radius
\be
\RB &=& \left[\frac{ {\mathbb{E}}({\mathbf{I}}_{\mathsf{LOS}})}{\lambda \pi}\right]^{\frac{1}{2}}.
\ee 
The plot of $\RB$ as a function of $\rout$ for various density of interferers is shown in Fig. \ref{fig:RoutVsRB} where we assume $W=0.3 ~\mathrm{m}$. From Fig. \ref{fig:RoutVsRB}, it is evident that for larger interferer density $\lambda$, the LOS ball radius $\RB$ quickly becomes insensitive to the variation of $\rout$. 
In the limit as $\rout \rightarrow \infty$, the expression for $\RB$ simplifies further as
\begin{eqnarray}
\RB &=& \frac{\sqrt{2}}{\lambda W} e ^{-\lambda \pi \frac{W^2}{8}}.
\end{eqnarray}
With this equivalent LOS ball assumption, $\pbr$ is a step function with a step up at distance $\RB$. This means that the interferers at a distance less than $\RB$ are \textit{strong} and those at distances greater than $\RB$ (i.e., falling outside the LOS ball) from the reference receiver are \textit{weak} and their combined signal power is treated as adding to the thermal noise variance. This simplifies the analysis considerably as shown later. The validity of this assumption is shown in Section \ref{sec:Sim_results}. 

\begin{figure}
\centering
\includegraphics[totalheight=2.6in, width=2.8in]{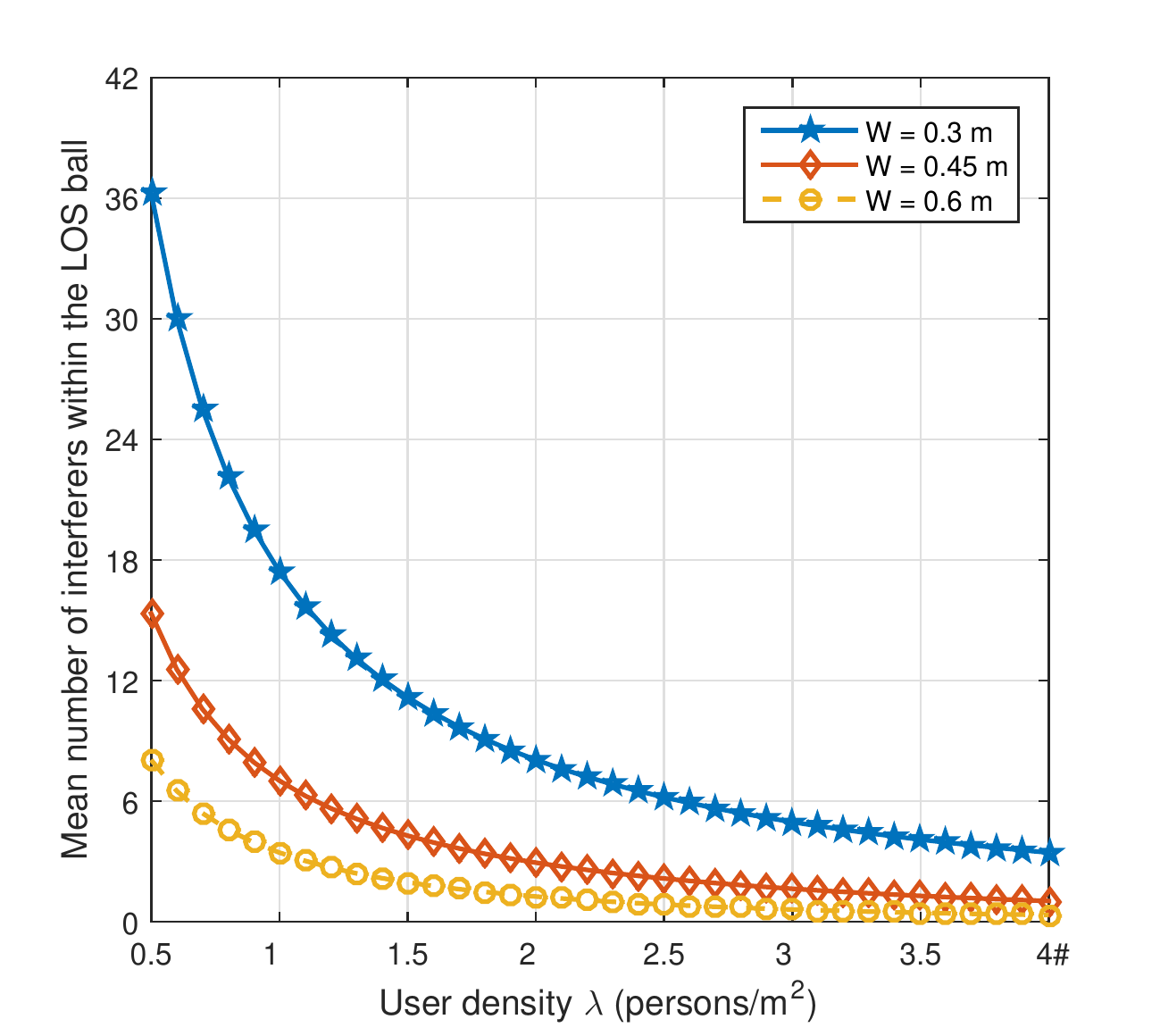}
\vspace{-0.1in}
\caption{Plot showing the mean number of interferers within the LOS versus the user density $\lambda$ for various values of human body blockage width $W$.}     
\label{fig:EXVslambda}        
\end{figure}

To better understand the impact of network density and human body blockage, the expected number of interferers within the LOS ball as a function of the user density $\lambda$ is shown in Fig. \ref{fig:EXVslambda} for various values of $W$. It is seen that unlike conventional networks, an increase in the interferer density $\lambda$ does not increase the density of the strong interferers. The mean number of strong interferers reduces for larger user density because both interferers and blockages are associated with users in mmWave wearable networks. So increasing interferer density $\lambda$ also increases the effect of blockages. Further, higher values for $W$ result in smaller mean number of strong interference in accordance to this intuition, as shown in Fig. \ref{fig:EXVslambda}.

Next we define the discrete random variable $I_i$ for $i \in \BPhi \cap \mathcal{B}(0, \RB)$, as follows. This is the random variable denoting the relative transmit power gain from $X_i$ that takes into account the random transmission probability $\pt$, and the random orientation of its antenna main-lobe.
\begin{eqnarray}
   I_i
   & = &
   \begin{cases}
        0 & \mbox{w. p. $(1-\pt)$} \\
        \Gt & \mbox{w. p. $\pt \left( \frac{\thetat}{2 \pi} \right)$} \\
        \gt & \mbox{w. p. $\pt \left( 1 - \frac{\thetat}{2 \pi} \right)$}
   \end{cases}. 
\end{eqnarray} The normalized power gain at the receiver for the interference signal from the $i^{\mathsf{th}}$ interferer $X_i$ depends on whether or not it falls in the main-lobe beam of the reference receiver. We defined the normalized power gain as \cite{mmWave:2015}
\begin{eqnarray}
   \Omega_i
   & = &
   \begin{cases}
      \frac{P_i}{P_0} \Gr R_i^{-\alpha_i} & \mbox{ if $-\frac{\thetar}{2} \leq \phi_i - \phi_0 \leq \frac{\thetar}{2}$} \\
      \frac{P_i}{P_0} \gr R_i^{-\alpha_i}& \mbox{ otherwise }
   \end{cases},     \label{Equation:Omegai}
\end{eqnarray} where $\alpha_i = \alphaL$ if $R_i \leq \RB$, and $=\alphaN$ if $R_i > \RB$. The random variable $R_i$, $i \in \BPhi \cap \mathcal{B}(0, \RB)$ in \eqref{Equation:Omegai} is independently and identically distributed (i.i.d) with probability density function (pdf)
\be
f_{R_i}(r) = \frac{2 r}{\RB^2},~ 0\leq r \leq \RB. \label{Equation:iidRi}
\ee Similarly, $R_i$, $i \in \BPhi \setminus \mathcal{B}(0, \RB)$ is also i.i.d with pdf
\be
f_{R_i}(r) = \frac{2 r}{\rout^2-\RB^2},~ \RB\leq r \leq \rout. \label{Equation:iidRi_NLOS}
\ee Defining $\IPhi = \sum_{i \in \BPhi \cap \mathcal{B}(0, \RB)} I_{i} h_{i} \Omega_i$, the SINR can be written as
\be
\gamma & = &
    \frac{ \Gr \Gt h_0 R_0 ^{-\alphaL}}
    {\displaystyle \sigma^2_{\mathsf{noise}} + \bar{\sigma^{2}}_{\mathsf{NLOS}} + \IPhi }. \label{Equation:SINR}
\ee In \eqref{Equation:SINR}, the quantity $\bar{\sigma^{2}}_{\mathsf{NLOS}}$ is the mean interference power due to the interferers lying outside the LOS ball. This is evaluated next.

\subsection{Evaluation of mean power due to weak interferers} \label{ssec_NLOS}
The average power of the interference outside the LOS ball is given by
\be
\nonumber \bar{\sigma^{2}}_{\mathsf{NLOS}} \hspace{-0.15in} &=&  \hspace{-0.1in}\mathbb{E} \hspace{-0.03in} \left[ \sum _{i \in \BPhi \setminus \mathcal{B}(0, \RB)} I_i h_i \Omega_i \right] \\
\hspace{-0.15in} &=& \hspace{-0.1in}\mathbb{E}\hspace{-0.03in}\left[ \sum _{i \in \BPhi \setminus \mathcal{B}(0, \RB)} \hspace{-0.28in} \pt \Gt \Omega_i \frac{\thetat}{2\pi} + \pt \hspace{-0.03in}\left(\hspace{-0.03in}1-\frac{\thetat}{2\pi}\hspace{-0.03in}\right)\hspace{-0.03in} \gt \Omega_i \right]\hspace{-0.05in}, \label{Equation:lim_NLOS}
\ee where we average over the fading $h_i$ and random transmission activity $I_i$. The number of interferers in the annular region $\mathcal{B}(0, \rout) \setminus \mathcal{B}(0, \RB)$ is Poisson distribution with mean $\lambda \mathcal{A}'$, where $\mathcal{A}' = \pi(\rout^2-\RB^2)$. Due to the circular symmetry of the annulus, the number of interferers ${K}'$ falling in the main-lobe of the receiver is Poisson distributed with mean $\frac{\thetar}{2\pi} \lambda \mathcal{A}'$. Similarly, the number of interferers ${K}''$ in the side-lobe of the receiver is also Poisson distributed with mean $\left(1-\frac{\thetar}{2\pi}\right) \lambda \mathcal{A}'$. Substituting $\Omega_i$ from \eqref{Equation:Omegai} and the pdf of $R_i$ per \eqref{Equation:iidRi_NLOS}, and noting that $|i \in \BPhi \setminus \mathcal{B}(0, \RB)| = {K}' + {K}''$, the RHS of \eqref{Equation:lim_NLOS} can be written as
\be
\nonumber\hspace{-0.1in} &&\hspace{-0.05in}\pt\hspace{-0.08in}\overset{~~\rout}{\underset{\hspace{-0.05in}\RB}{\int}} \hspace{-0.07in} r^{-\alphaN} \frac{2\pi r}{\mathcal{A}'} \mathsf{d}r\hspace{-0.03in}\left(\frac{\thetat}{2\pi}\Gt \hspace{-0.03in}+ \hspace{-0.03in}\left(1- \frac{\thetat}{2\pi}\right)\hspace{-0.03in}\gt\hspace{-0.03in}\right)\hspace{-0.03in} \mathbb{E}\left[ {{K}'} \Gr + {{K}''}\gr\right], 
\ee so that 
\be
\nonumber \bar{\sigma^{2}}_{\mathsf{NLOS}} \hspace{-0.05in}&=& \hspace{-0.05in} \frac{2\pi \lambda (\RB^{2-\alphaN} - \rout^{2-\alphaN}) \pt }{\alphaN-2}\left( \frac{\thetat}{2\pi}\Gt + (1-\frac{\thetat}{2\pi})\gt \right) \\
&& ~~~~~~~~~~~~\times \left(\frac{\thetar}{2\pi} \Gr + (1-\frac{\thetar}{2\pi})\gr \right). \label{Equation:NLOS_avg}
\ee By using
\be
\textbf{q}
   & \hspace{-0.1in} = \left[\hspace{-0.05in}\begin{array}{c}
   \frac{\thetat \thetar}{4\pi^2} \\ 
   \left( 1 - \frac{\thetat }{2\pi} \right)\frac{\thetar}{2\pi} \\ 
   \frac{\thetat }{2\pi} \left( 1-\frac{\thetar}{2\pi}\right) \\ 
   \left( 1- \frac{\thetat }{2\pi} \right)\left( 1-\frac{\thetar}{2\pi}\right)
   \end{array} \hspace{-0.05in}\right]
        ~ \mathrm{and}~
\textbf{G}
   & \hspace{-0.1in}= 
   \left[\begin{array}{c}
        \Gt\Gr \\
        \gt\Gr \\
        \Gt\gr \\
        \gt\gr
   \end{array}\right],
\ee \eqref{Equation:NLOS_avg} can be written compactly as
\be
\bar{\sigma^{2}}_{\mathsf{NLOS}} \hspace{-0.05in}&=& \hspace{-0.05in} \frac{2\pi \lambda (\RB^{2-\alphaN} - \rout^{2-\alphaN}) \pt }{\alphaN-2} {\textbf{q}}^{\mathrm{T}}\textbf{G}. \label{Equation:NLOS_avg_compact}
\ee

\subsection{Coverage probability}
\label{ssec_cov}
The coverage probability ${P}_{\mathsf{c}}(\beta)$ is defined as the complementary cumulative distribution function (CCDF) of the SINR evaluated at a threshold $\beta$
\begin{eqnarray}
P_{\mathsf{c}}(\beta) = \mathbb{E}_{\BPhi}\left[\mathbb{P}\left[\gamma > \beta | \BPhi \right]\right], \label{Equation:Coverage}
\end{eqnarray} where $\mathbb{P}\left[\gamma > \beta ~|~ \BPhi \right]$ is the CCDF of SINR evaluated at $\beta$ for a given realization of $\BPhi$. Substituting \eqref{Equation:SINR} into \eqref{Equation:Coverage} defining $\sigma^2 = \sigma^2_{\mathsf{noise}} + \bar{\sigma^{2}}_{\mathsf{NLOS}}$, and rearranging leads to
\begin{eqnarray}
P_{\mathsf{c}}(\beta) & = & \mathbb{E}_{\BPhi}\left[\mathbb{P}\left[ \beta^{-1} \frac{\Gt \Gr}{R_0^{\alphaL}} h_0  > \sigma^2 + \IPhi \right] \right]. \label{Equation:SINR_exp}
\end{eqnarray}
For a normalized gamma distributed random variable with (integer) parameter $m$, the CDF evaluated at a point $x$ can be tightly lower bounded by $\left( 1-e^{-m \tilde{m} x}\right)^m$, where $\tilde{m} = {(m!)^{\frac{-1}{m}}}$ \cite{Alzer:1997}. Using this result in \eqref{Equation:SINR_exp} with $\tilde{\beta} = \frac{ \beta R_0^{\alphaL}}{\Gt \Gr}$, 
\begin{eqnarray}
P_{\mathsf{c}}(\beta)\leq 1- \mathbb{E}_{\BPhi} \left[\left( 1 - e^{-m \tilde{m} \tilde{\beta}(\sigma^2 + \IPhi)}\right)^m \right]. \label{Equation:SINR_approx}
\end{eqnarray}
Using the binomial theorem, \eqref{Equation:SINR_approx} can be expanded and
\begin{eqnarray}
P_{\mathsf{c}}(\beta)\leq \sum_{\ell = 1}^{m}\binom {m}{\ell} (-1)^{\ell + 1}e^{-\ell m \tilde{m} \tilde{\beta}\sigma^2} \mathbb{E}_{\BPhi}\left[ e^{-\ell m \tilde{m} \tilde{\beta}\IPhi} \right]. \label{Equation:SINR_binom}
\end{eqnarray} The expectation in \eqref{Equation:SINR_binom}, simplifies further as
\be
\nonumber \mathbb{E}_{\BPhi}\left[ e^{-\ell m \tilde{m} \tilde{\beta}\IPhi} \right] \hspace{-0.1in}&=& \hspace{-0.1in} \mathbb{E}_{\Omega_i,h_i,I_i}\hspace{-0.07in}\left[ e^{-\ell m \tilde{m} \tilde{\beta}\sum_{i \in \BPhi \cap \mathcal{B}(0, \RB)} I_{i} h_{i} \Omega_i} \right] \\
\hspace{-0.1in}& = &\hspace{-0.1in} \mathbb{E}_{\Omega_i}\hspace{-0.05in}\left[ \hspace{-0.03in}\prod_{i \in \BPhi \cap \mathcal{B}(0, \RB)}\hspace{-0.3in} \mathbb{E}_{h_i,I_i}\hspace{-0.05in} \left[ e^{-\ell m \tilde{m} \tilde{\beta} I_{i} h_{i} \Omega_i} \right] \hspace{-0.04in}\right]\hspace{-0.04in}, \label{Equation:expanden_PPPform}
\ee where the last step uses the fact that $\{h_i\}$ and $\{I_i\}$ are independent. Noting that $h_i$ is a Gamma random variable, the term $\mathbb{E}_{h_i}\left[ e^{-\ell m \tilde{m} \tilde{\beta} I_{i} h_{i} \Omega_i} \right]$ is its moment generating function which evaluates to
\be
\mathbb{E}_{h_i}\left[ e^{-\ell m \tilde{m} \tilde{\beta} I_{i} h_{i} \Omega_i} \right] &=& \left( 1 + \ell \tilde{m} \tilde{\beta} I_{i} \Omega_i  \right)^{-m},
\ee so that
\be
\nonumber \mathbb{E}_{h_i,I_i} \hspace{-0.05in}\left[ e^{-\ell m \tilde{m} \tilde{\beta} I_{i} h_{i} \Omega_i} \right] \hspace{-0.12in}&=& \hspace{-0.12in}(1-\pt) + \pt\hspace{-0.05in}\left[ \frac{\thetat}{2\pi} (1+\ell \tilde{m} \tilde{\beta} \Gt \Omega_i)^{-m} \right. \\
\hspace{-0.15in} && \hspace{-0.15in} +\left. \left(\hspace{-0.03in}1-\frac{\thetat}{2\pi}\hspace{-0.03in}\right)\hspace{-0.03in} (1+\ell \tilde{m} \tilde{\beta} \gt \Omega_i)^{-m}\right].
\ee The number of interferers ${K}$ that are within $\mathcal{B}(0, \RB)$ is Poisson distributed with mean $\lambda \pi \RB ^2$. Given ${{K}}$, the number of interfering transmitters $\tilde{{K}}$ falling within the receiver antenna main-lobe is a Binomial random variable with parameter $\frac{\thetar}{2\pi}$. Using \eqref{Equation:Omegai} and denoting $\mathcal{T}(\mathfrak{g},R)$ as 
\be
\nonumber \mathcal{T}(\mathfrak{g},R) \hspace{-0.1in}&=&\hspace{-0.1in} (1-\pt) + \pt\left[ \frac{\thetat}{2\pi} (1+\ell \tilde{m} \tilde{\beta} \Gt \mathfrak{g} R^{-\alphaL})^{-m} \right. \\
\hspace{-0.1in}&&\hspace{-0.1in} ~~+\left. \left(1-\frac{\thetat}{2\pi}\right) \hspace{-0.05in}(1+\ell \tilde{m} \tilde{\beta} \gt \mathfrak{g} R^{-\alphaL})^{-m} \right]  \\
\text{and }
\mathbb{P}_{\tilde{K}}(n) &=& \binom {{K}}{n} \left( \frac{\thetar}{2\pi}\right)^n \left(1- \frac{\thetar}{2\pi}\right)^{{{K}}-n},
\ee the RHS of \eqref{Equation:expanden_PPPform} can be expanded as follows, 
\be
\nonumber
\hspace{-0.2in}&&\hspace{-0.3in}\mathbb{E}_{\BPhi}\left[ e^{-\ell m \tilde{m} \tilde{\beta}\IPhi} \right] \\ 
\nonumber \hspace{-0.2in}&=& \hspace{-0.1in} \mathbb{E}_{{K}} \hspace{-0.05in}\left[ \mathbb{E}_{R_i}\hspace{-0.05in}\left[ \sum_{n=0}^{{K}} \mathbb{P}_{\tilde{K}}(n)(\mathcal{T}(\Gr,R_i))^n   (\mathcal{T}(\gr,R_i))^{{{K}}-n} \right]\hspace{-0.02in}\right] \label{Equation:SINR_binomexp}\\
\hspace{-0.2in}&=&\hspace{-0.1in}\mathbb{E}_{{K}} \hspace{-0.05in} \left[ \mathbb{E}_{R_i}\hspace{-0.07in}\left[ \hspace{-0.02in}\left(\frac{\thetar}{2\pi}\hspace{-0.03in} \mathcal{T}(\Gr,R_i) + \hspace{-0.02in}\left( 1- \frac{\thetar}{2\pi} \right)\hspace{-0.02in}\mathcal{T}(\gr,R_i)\hspace{-0.03in}\right)^{{K}} \right]\hspace{-0.02in} \right]\hspace{-0.05in}. \label{Equation:expKexpR}
\ee Note that for a given ${{K}}$, $\{R_i\}, i = 1,..., {{K}}$ are i.i.d with pdf as given in \eqref{Equation:iidRi}. So \eqref{Equation:expKexpR} becomes
\be
\nonumber \hspace{-0.15in} &=& \hspace{-0.1in}
\mathbb{E}_{{K}} \hspace{-0.05in}\left[\hspace{-0.04in}  \left(\mathbb{E}_{R_i}\hspace{-0.03in}\left[ \hspace{-0.03in}\frac{\thetar}{2\pi} \mathcal{T}(\Gr,R_i) + \left( 1- \frac{\thetar}{2\pi} \right)\mathcal{T}(\gr,R_i)\right]\right)^{{K}}  \right]\\ 
\nonumber \hspace{-0.15in} &=& \hspace{-0.1in}\mathbb{E}_{{K}} \hspace{-0.06in}\left[ \hspace{-0.09in} \overset{\hspace{0.1in}\RB}{\underset{\hspace{-0.1in}0}{\int}} \hspace{-0.05in}\frac{2r}{R_B^2}\left(1- \pt +\hspace{-0.05in} \sum_{i=1}^4 \pt q_i (1+\ell \tilde{m} \tilde{\beta} G_i r^{-\alphaL})^{-m}\hspace{-0.03in}\right)\hspace{-0.03in}\mathsf{d}r\hspace{-0.03in}\right]^{{K}} \\
 \hspace{-0.15in} &=& \hspace{-0.1in} \exp\hspace{-0.04in}\left(\hspace{-0.06in} -\lambda \pi \pt \hspace{-0.06in}\left(\hspace{-0.06in}\RB^2 \hspace{-0.05in}- 2\sum_{i=1}^4 \hspace{-0.05in} q_i \hspace{-0.1in}\overset{\hspace{0.05in}\RB}{\underset{\hspace{-0.05in}0}{\int}}\hspace{-0.1in}(1+\ell \tilde{m} \tilde{\beta} G_i r^{-\alphaL})^{-m} r \mathsf{d}r\hspace{-0.05in}\right)\hspace{-0.05in}\right)\hspace{-0.05in}. \label{Equation:FinalInt}
\ee The $i^{\mathrm{th}}$ term of $\mathbf{q}$ and $\mathbf{G}$ are denoted as $q_i$ and $G_i$, respectively in \eqref{Equation:FinalInt}.

Given the CCDF of SINR, the distribution of spectral efficiency $\eta$ can be evaluated, using which ergodic spectral efficiency can be computed as $\mathbb{E}\left[\eta \right]$.

\section{Numerical Results}
\label{sec:Sim_results} 
\begin{figure}
\centering
\includegraphics[totalheight=2.6in, width=2.8in]{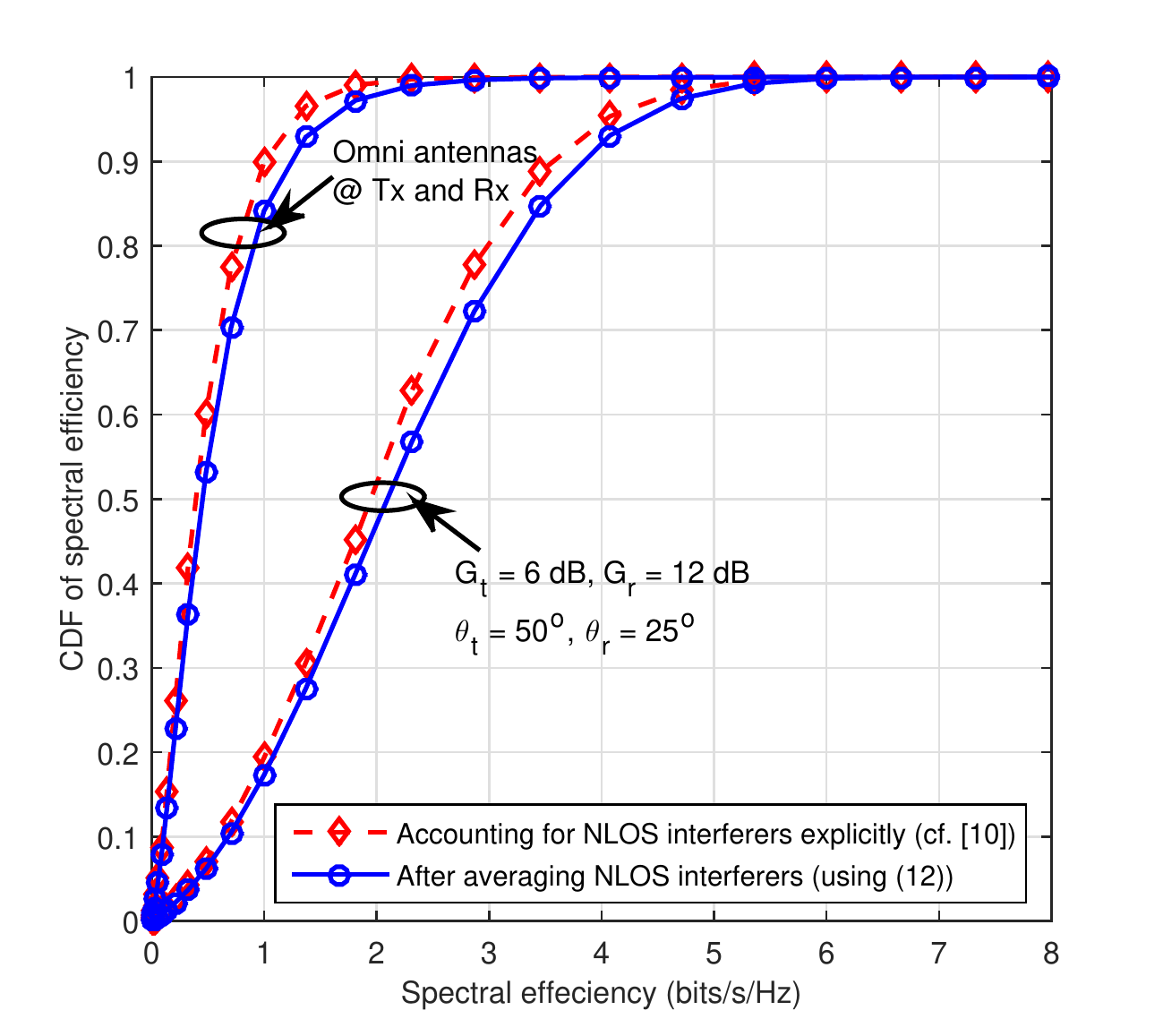}
\vspace{-0.1in}
\caption{Plot showing the CDF of spectral efficiency with and without the averaging assumption for the interference from the NLOS interferers for different antenna configurations when the transmission probability $\pt = 1$, $\lambda = 3 {\mathrm{m}}^{-2}$ and $W = 0.3 ~\mathrm{m}$.}
\label{fig:Spectral_NLOSw_wo}        
\end{figure}

\renewcommand*{\thefootnote}{\fnsymbol{footnote}}
In this section, we first show the validity of our assumption of considering only the strong interferers within the LOS ball and treating the weak interferers as adding to the thermal noise variance. We compare the plots of ergodic efficiency distribution of the actual network without averaging the weak interferers that lie outside $\mathcal{B}(0, \RB)$ (see \cite{mmWave_arXiv:2015} for details) against the results obtained via the formulation in Section \ref{ssec_NLOS}. This is shown in Fig. \ref{fig:Spectral_NLOSw_wo}. We conclude that it is reasonable to capture the NLOS mean interference as a factor that adds to the thermal noise floor.  

\begin{figure}
\centering
\includegraphics[totalheight=2.6in, width=2.8in]{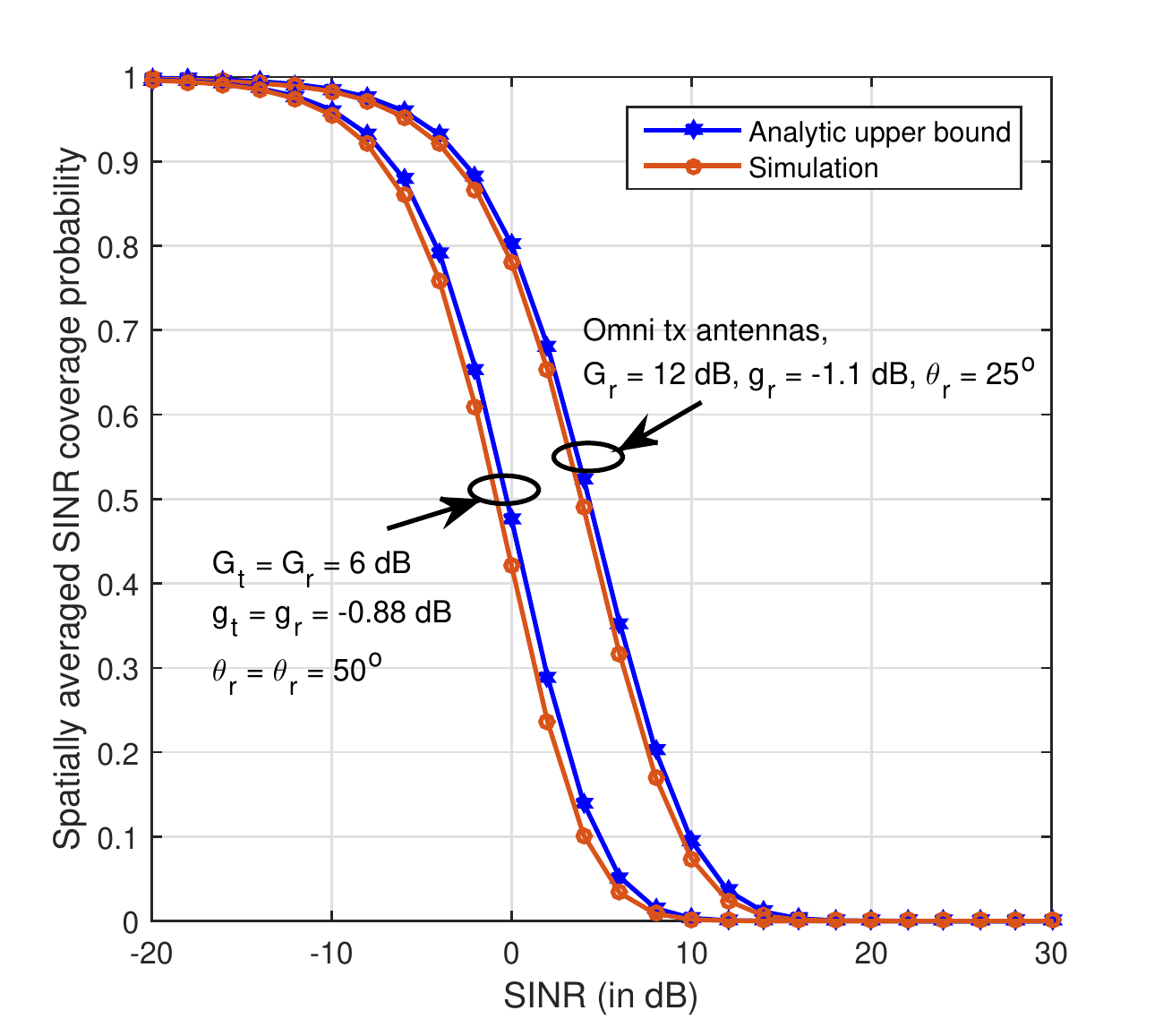}
\vspace{-0.1in}
\caption{Plot showing the SINR distribution obtained through simulation and analytic expression for different antenna configurations when the transmission probability $\pt = 0.8$ .}     
\label{fig:SINRdistbn}        
\end{figure}

\begin{figure}
\centering
\includegraphics[totalheight=2.6in, width=2.8in]{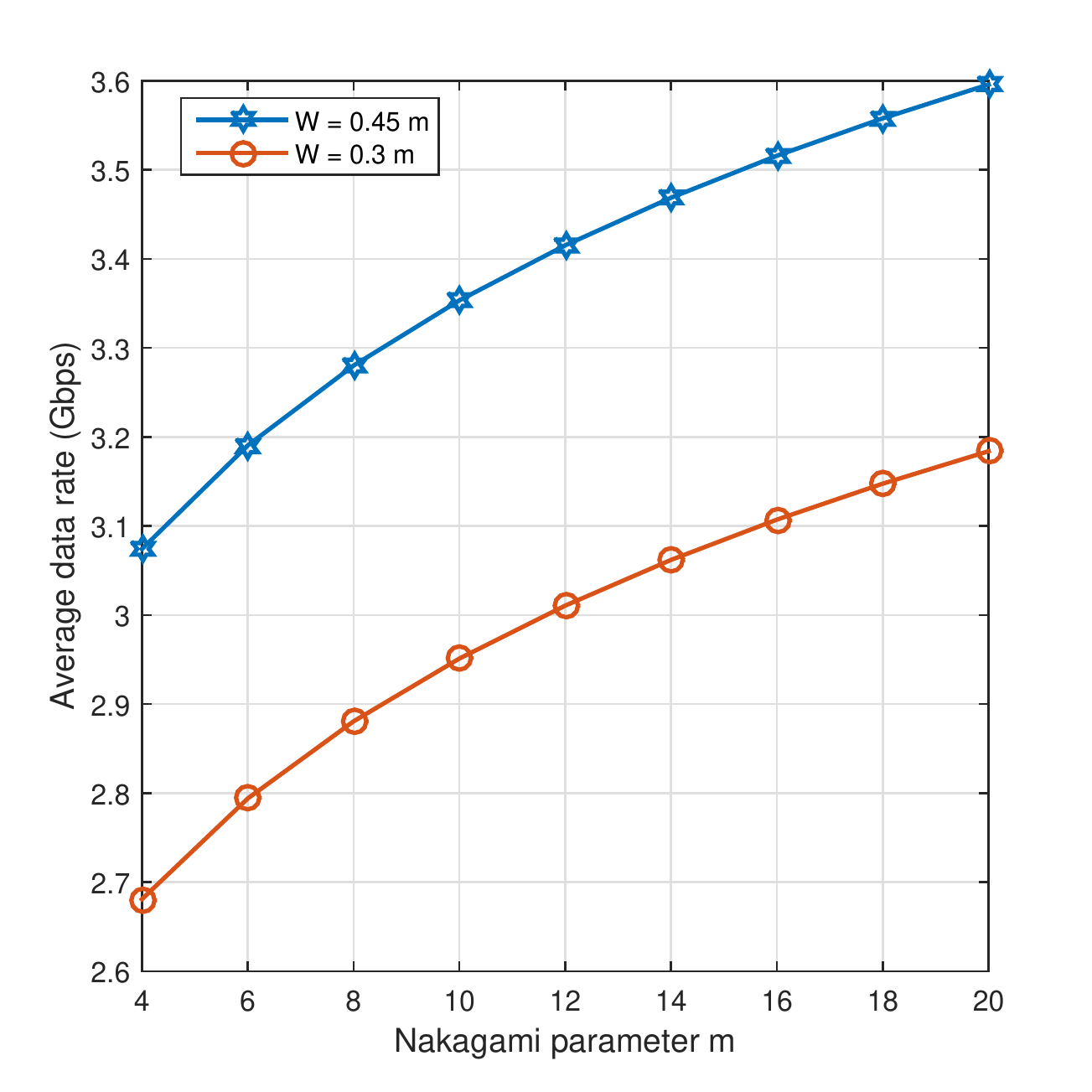}
\vspace{-0.1in}
\caption{Plot showing the effect of the Nakagami fade gain parameter $m$ for the link between the strong interferer and the reference receiver for $\lambda = 2~{\mathrm{m}}^{-2}$.}     
\label{fig:specEvs_m}        
\end{figure}

Fig. \ref{fig:SINRdistbn} shows the analytic results against that obtained via Monte Carlo simulation of random placement of strong interferers within the LOS ball for the case when $W=0.3~\mathrm{m}$, $\lambda = 3~{\mathrm{m}}^{-2}$, and $\pt = 0.8$ for two different antenna configurations. This shows that the upper bound for SINR CCDF used in \eqref{Equation:SINR_approx} is very tight and also validates the closed-form expressions derived in Section \ref{ssec_cov}. 
The effect of the Nakagami parameter $m$ on the average data rate is shown in Fig. \ref{fig:specEvs_m} for two different interferer densities for $\Gt = \Gr = 6~\text{dB}$, $\gt = \gr = -0.88~\text{dB}$, $\thetat = \thetar = 50^o$, and $\pt = 1$. Increasing $m$ results in the LOS links being more like AWGN, and the system performance is seen to improve similar to that observed in free space point-to-point communication. 

\section{Conclusion}
Modeling wearable networks involves modeling the human blockages and potential interferers that are both associated with users. We show that assuming the blockages and the interferers are drawn from a PPP in an infinite two-dimensional plane and explicitly accounting for the interferers that lie within a LOS ball while treating the NLOS interference that lie outside the ball as adding to noise, we can closely approximate the realistic case of finite number of interferers in a finite network region when the user density is high. We also develop closed-form expressions for spatially averaged SINR coverage that incorporates antenna parameters and density. The limiting case of AWGN channel for the LOS links is also studied to evaluate the sensitivity of the performance to strong LOS interference within close proximity. 

\bibliographystyle{ieeetr}

\end{document}